# Tuning Spin Dynamics and Localization Near the Metal-Insulator Transition in Fe/GaAs heterostructures

Yu-Sheng Ou[1], N. J. Harmon[2], Patrick Odenthal[3], R. K. Kawakami[1,3], M. E. Flatté[2], E. Johnston-Halperin[1†]

[1]*Department of Physics, The Ohio State University, Columbus, OH 43210-1117, USA*

[2]*Department of Physics & Astronomy, The University of Iowa, Iowa City, IA 52242-1479, USA*

[3]*Department of Physics & Astronomy, University of California, Riverside, CA 92521, USA*

Abstract

We present a simultaneous investigation of coherent spin dynamics in both localized and itinerant carriers in Fe/GaAs heterostructures using ultrafast and spin-resolved pump-probe spectroscopy. We find that for excitation densities that push the transient Fermi energy of photocarriers above the mobility edge there exist two distinct precession frequencies in the observed spin dynamics, allowing us to simultaneously monitor both localized and itinerant states. For low magnetic fields (below 3 T) the beat frequency between these two excitations evolves linearly, indicating that the nuclear polarization is saturated almost immediately and that the hyperfine coupling to these two states is comparable, despite the 100x enhancement in nuclear polarization provided by the presence of the Fe layer. At higher magnetic fields (above 3 T) the Zeeman energy drives reentrant localization of the photocarriers. Subtracting the constant hyperfine contribution from both sets of data allows us to extract the Lande g-factor for each state and estimate their energy relative to the bottom of the conduction band, yielding -2.16 meV and 17 meV for localized and itinerant states, respectively. This work advances our fundamental understanding of spin-spin interactions between electron and nuclear spin species, as well as between localized and itinerant electronics states, and therefore has implications for future work in both spintronics and quantum information/computation.

**I. Introduction**

Defects play a central role in the developing fields of spintronics and quantum information, whether they are viewed as a loss channel for spin coherence in spin transport[1–8] or are themselves systems of interest for quantum computation[9,10] or quantum communication[11–14]. Many of these phenomena and proposed applications rely sensitively on the interplay between these defect states and itinerant carriers. For example, early work focused on maximizing the spin lifetime in semiconducting transport channels revealed that the lifetime is maximal in the vicinity of the metal-insulator transition[8,15–18] while many schemes for solid-state quantum computing rely on using conduction band spins to coherently bridge between defect states[9,10]. Analysis of these interactions is challenging due to the complexity of the multiple channels for spin coupling and interaction, ranging from the variety of ways defect scattering can interact with spin-orbit coupling (such as D'yakanov-Perel spin relaxation) to differences in hyperfine coupling due to potential differences in wave function overlap between localized and itinerant electrons[8,19,20]. However, despite the longstanding importance of these interactions, there have been relatively few studies to date that have been able to simultaneously explore the coherent spin dynamics of both localized and itinerant states within the same experiment[21,22].

Here we present a study of coherent spin dynamics in Fe/GaAs heterostructures where we tune the effective Fermi energy across the mobility edge by systematically tuning the density of photocarriers created by optical excitation. This is done using a time-resolved pump-probe technique that allows for the monitoring of the coherent spin dynamics in the sample, including both localized and itinerant electrons, and therefore reveals the differences, if any, between these two populations in the same sample and under the same experimental conditions. In addition, as we have previously demonstrated[2,23,24], the inclusion of the Fe layer significantly enhances the nuclear spin polarization in these samples. This in turn dramatically amplifies the hyperfine interaction (roughly 100 times stronger than in isolated GaAs epilayers), allowing us to sensitively probe for any small differences in the hyperfine coupling between these two states. We find that for excitation densities that push the effective Fermi energy above the mobility edge there exist two distinct precession frequencies in the observed spin dynamics, allowing us to simultaneously monitor both localized and itinerant states. For low magnetic fields (below 3 T) the beat frequency between these two excitations evolves linearly, indicating that the nuclear polarization is saturated almost immediately and that the hyperfine coupling to these two states is comparable. At higher magnetic fields (above 3 T) the Zeeman energy drives reentrant localization of the photocarriers. Subtracting the constant hyperfine contribution from both sets of data allows us to extract the Lande g-factor for each state and estimate their energy relative to the bottom of the conduction band, yielding -2.7 meV and 17 meV for localized and itinerant states, respectively.

**II. Sample Synthesis and Ultrafast Spin Probes**

The samples studied here are prepared via metal-organic chemical vapor deposition (MOCVD) and molecular beam epitaxy (MBE) in accordance with Refs. 2 and 24, with a layer structure of 10 nm MgO/10 nm Fe/0.2 nm MgO/120 nm Si-doped n-type GaAs ($n = 7\times10^{16}/cm^3$)/400 nm $In_{0.5}Ga_{0.5}P$/$n^+$-type GaAs substrate. The Fe layer serves to enhance nuclear polarization via interfacial exchange coupling[2,25,26] and the thickness of the MgO layer between Fe and n-GaAs epitaxy layer is selected to optimize the exchange coupling between Fe and n-GaAs while preventing interfacial intermixing which can give rise to an intermetallic FeGa phase[24]. The samples are mounted face-down on 100 µm thick sapphire wafers so that the $n^+$-GaAs substrates can be removed by selective wet etching using the $In_{0.5}Ga_{0.5}P$ layer as a chemically-selective etch stop[27].

Since the carrier concentration in the samples studied in this work is comparable to the room temperature metal insulator transition (MIT) in GaAs ($n_{MIT} \sim 3\times10^{16}/cm^3$)[8], we assume that the Si donor band (DB) hybridizes with the conduction band (CB), resulting in both occupied and unoccupied localized states at low energy and itinerant states at higher energy, as shown in Fig. 1 (a). This further suggests that the Fermi level, $E_F$, is located near the boundary between localized and delocalized states (direct measurement of the MIT in these samples is precluded by parasitic conduction in the InGaP layer).

Time resolved Faraday rotation (TRFR) spectroscopy is employed to study the GaAs electron spin dynamics. The technique is briefly summarized as follows: a circularly-polarized (CP) pump pulse, tuned to the band edge of GaAs ($E_{pump}$= 817 nm or 1.517 eV), excites a spin ensemble in GaAs along its propagation direction and with an energy distribution given by the spectral width of the laser (approximately 10 meV) as shown in Fig. 1(a). For comparison, Fig. 1(b) shows the effect of increasing the intensity of the pump beam without changing its spectral position or width. The increased photon density translates to an increase in photo-carrier density and a consequent increase in both majority and minority spin populations. In either case, the photoexcited spin ensemble relaxes to the bottom of the conduction band within a picosecond[28], and starts to precess coherently in the presence of a transverse magnetic field, $B_{tot}$. The Faraday rotation angle ($\theta_{FR}$) of a much weaker time-delayed linearly-polarized (LP) probe pulse ($E_{probe}$) directly measures the instantaneous component of this spin ensemble along its propagation direction. By systematically varying the delay time ($\Delta t$) between pump and probe pulse, the temporal evolution of the coherent photoexcited spin ensemble is revealed. In this work, laser pulses of 130-fs duration and 76 MHz repetition rate are generated by a mode-locked Ti: sapphire laser with central wavelength at 817 nm, and are split into pump and probe pulse trains whose power ratio is always kept above 10 with a time-averaged probe power of 0.31 mW (power density of 4 W/cm$^2$). All the TRFR measurements in this work are taken at temperature $T = 40$ K in a liquid Helium magneto-optical cryostat.

**III. Identification of Localized and Itinerant Carriers**

Figure 1 (c) shows a typical TRFR time scan taken on a Fe/MgO/GaAs heterostructure with an in-plane applied field $B_{app} = 5$ T and with a pump power density $I = 48$ W/cm$^2$. This time trace can be described by the following equation:

$$\theta_{FR}(\Delta t) = A \left( e^{-\left(\Delta t / T_2^*\right)} + N_0 e^{-\left(\Delta t / T_h\right)} \right) \cos(2\pi f \Delta t + \theta), \quad (1)$$

where $A$ is the maximal Faraday angle and $N_0$ is the ratio of photoexcited to equilibrium electrons ($N_0 = N_{ex}/n$) at $\Delta t = 0$, $T_2^*$ is the inhomogeneous dephasing time of the photoexcited spin ensemble, $T_h$ is the hole carrier recombination time, $f = \dfrac{g_{eff} \mu_B B_{tot}}{h}$ is the Larmor precession frequency caused by the total magnetic field $B_{tot} = B_{app} + B_n$, where $B_{app}$ is the external applied field and $B_n$ represents a hyperfine-driven effective field from GaAs nuclei[29], and $\theta$ is the phase of spin precession. The two exponential terms reflect the fact that the photoexcited holes, while not directly detected due to their rapid spin relaxation, do act to dephase the electron ensemble through the Bir-Aranov-Pikus mechanism until they recombine (typically in less than 100 ps). However, when the pump intensity increases to $I = 241$ W/cm$^2$, corresponding to a photocarrier density of $8.8 \times 10^{16}$ cm$^{-3}$, we find that this single-frequency fitting function no longer provides a good description of the data (Fig. 1(d)).

A quick and fitting-function-independent approach to determine the possible origin of this discrepancy is to compare the fast Fourier transformation (FFT) power spectra for low- and high-intensity regimes (Fig. 1 (e)). The top panel of Fig. 1(e) shows the FFT for the low power data shown in Fig. 1(c), revealing a single peak whose width is consistent with the spin lifetime extracted from a fit of the time domain data to Eq. (1) ($T_2^* = 0.308$ ns). In contrast, the bottom panel of Fig. 1(e) shows the FFT of the high power data shown in Fig. 1(d), and in addition to the narrow peak seen in the top panel a broad peak at lower frequency is revealed. This suggests that the origin of the failure of Eq. (1) in the high power regime is due to the emergence of a second precession with lower frequency and shorter lifetime.

To quantitatively explore this behavior, we add a second exponentially damped cosine function to eq. (1) and remove the exponential term that is related to the hole recombination time as it becomes commingled with the spin lifetime of the low-frequency component,

$$\theta_{FR}(\Delta t) = A_H e^{\left(-\Delta t / (T_2^*)_H\right)} \cos(2\pi f_H \Delta t + \theta_H) + A_L e^{\left(-\Delta t / (T_2^*)_L\right)} \cos(2\pi f_L \Delta t + \theta_L), \quad (2)$$

where the first and second term represent the spin dynamics of the high and low frequency components, indicated by the subscripts $H$ and $L$, respectively. The time scale of second spin population is found comparable to that of the hole recombination observed in the low power regime (< 100 ps), but its amplitude is orders of magnitude larger. As a result, for $\Delta t <$ 100 ps, $\theta_{FR}$ is dominated by the second spin population, validating our decision to ignore the effect of recombination. The results of fitting the high power data with Eq. (2) can be seen in Fig. 1(f), and the quality of the fit ($\chi^2$) is reduced by an order of magnitude. The extracted values of $(T_2^*)_L$ and $f_L$ (21 ps and 26.41 GHz, respectively) are consistent with the values extracted from the FFT in Fig. 1(e), confirming the presence of a second precession frequency and consequently a second spin population in the sample that is only accessed by high pump fluence.

A potential explanation for this emergent state can be found from careful consideration of the schematic density of states (DOS) diagrams in Figs. 1(a) – (b). Note that while the center energy and energy width of the pump beam do not change, the fact that the ultrafast laser has a finite spectral width (~10 meV) means that as the intensity increases the high energy tail of the spectral distribution can continue to add carriers well above the Fermi energy. As a result, the transient Fermi energy of the photocarriers will continue to increase with increasing pump fluence. If the initial doping of the GaAs is below the mobility edge of the metal insulator transition (as is the case here) then this can transiently drive the Fermi energy above that mobility edge, creating a population of itinerant photocarriers in parallel with the localized carriers excited at low pump power. A second consequence of the spectral width of the ultrafast laser is that the TRKR will be sensitive to the spin dynamics of all states ($\Delta E_{LP}$ or $\Delta E_{HP}$) that fall within the spectral window of the probe laser pulse, i.e. it will simultaneously resolve the dynamics of both localized and itinerant carriers.

**IV. Coherent Spin Dynamics of Localized and Itinerant Carriers**

As reported previously, the presence of an Fe epilayer in these heterostructures serves to amplify the nuclear hyperfine coupling by roughly 100 times[2]. This sensitivity, combined with the ability to simultaneously monitor the coherent spin dynamics of both localized and itinerant carriers, makes these samples an excellent testbed for exploring the impact of localization length on the hyperfine interaction in solids.

Complementing the Fermi energy tuning demonstrated in Section III, here we vary the magnetic field from 1T to 6T at a variety of pump power densities ranging from 48 W/cm$^2$ to 455 W/cm$^2$. Figure 2 shows the results of these studies focusing on the extracted Larmor precession frequencies for both the high- and low-frequency components (Figs. 2 (a) – (d)) and the corresponding FFTs (Figs. 2(e) – (h)). As discussed previously, at I= 48 W/cm$^2$ there is only a single frequency component in both the time domain and frequency domain data (Figs. 2(a) and 2(e)), but at higher power densities a second low-frequency component emerges. In all cases the variation of frequency with applied magnetic field appears to be dominated by linear contributions. In this regime the nuclear polarization is expected to be saturated, as the fields applied are larger than the saturation field of Fe[2,24], and so we tentatively assign this linear dispersion to be due to the Zeeman interaction ($\hbar\Omega_L = g_{eff}\mu_B B_{app}$). A careful inspection of Figs. 2(b) – (d) suggests the presence of two distinct behaviors as a function of power.

Figure 3 explores this variation in more detail. Figure 3(a) provides a detailed power dependence at fixed applied field (5.000 T), and shows a clear threshold behavior wherein at the very lowest power there is only a single frequency component, but as the power density is increased from 60 - 121 W/cm$^2$ a sharp variation in frequency is observed for the low frequency component. Above this threshold the low-frequency component stabilizes for the remaining range of powers accessible to the experiment (151 W/cm$^2$ – 455 W/cm$^2$). This sharp threshold behavior is hard to understand in the context of simply filling carriers into the parabolic conduction band, and strongly supports the idea of a density of states that includes a mobility threshold with diverging behavior for localized and delocalized states. Further support for this model can be found in estimating the density of photocarriers as a function of pump fluence, taking into account both the absorption coefficient of GaAs and the carrier lifetime of this sample. The result reveals that the threshold in pump power density corresponds to the regime where the density of photocarriers becomes comparable to the density of native carriers supplied by the Si donors in the GaAs matrix ($7\times10^{16}$/cm$^3$), e.g. the regime where the density of photocarriers becomes sufficient to transiently perturb the Fermi energy to a significant degree.

This model can be further explored by considering the beat frequency between the high- and low-frequency components as a function of applied field. Figure 3(b) shows such data for a pump fluence of 108 W/cm$^2$ (red diamonds). Surprisingly, the beat frequency shows a maximum at a field of roughly 3 T, eventually returning to zero by 6 T. This non-monotonic behavior can be understood as the interaction between the g-factors of the localized and itinerant states (Fig. 3(c)). Since the localized states have a higher g-factor, as the applied field increases they will move completely below the relatively modest Fermi energy of the photocarriers, resulting in a reentrant localization of the entire ensemble in the high field regime. In a similar vein, the higher fluence curves show a similar crossover at slightly higher applied field, but as these higher densities of photocarriers exceed the total number of localized states in the system only a fraction of the ensemble becomes localized. Presumably, were higher fluences accessible without damaging the sample, then the linear regime of beat frequency would extend for the full field range as the relative number of localized and itinerant carriers would remain constant.

It is important to note that none of this discussion requires explicit consideration of the hyperfine coupling and associated effective field, $B_n$. This is quite surprising given the expected dependence of the Overhauser effect on the spatial distribution of the electron wavefunction and the myriad ways in which the localization length is modulated in these experiments. In fact, the beat frequency between the high- and low-frequency components extrapolates to 0 Hz at 0 T to within +/- 0.2 GHz, indicating that the effective field due to the hyperfine interaction is comparable for localized and delocalized carriers across all observed regimes. Further, the fact that neither frequency evolves with pump power for power density greater than 150.6 W/cm² (Fig. 3(a)) supports the assertion that the nuclear polarization is saturated throughout our experimental regime (if this were not the case the increase in photocarrier density would be expected to generate a commensurate increase or decrease in precession frequency[24,30] depending on the sign of effective nuclear field with respect to external applied field).

This lack of sensitivity of the hyperfine field to localization length or pump power allows for the extraction of an effective Lande g-factor ($g_{eff}$) as a function of pump power density (Fig. 3(c)). The evolution of the g-factor closely tracks the evolution of precession frequency, and we can use a $k \cdot p$ model[31] to determine the following energy dependence of $g$ factor in the conduction band:

$$g_{eff}(E) = -0.44 + 6.3 \times E(eV), \quad (3)$$

where -0.44 is the $g$ factor at the conduction minimum and $E$ is the excess energy from the conduction band minimum. This analysis gives $E = -2.16$ meV for the localized carriers and $E = 17$ meV for the itinerant carriers.

## V. Conclusions

We have developed a system that allows for the simultaneous investigation of coherent spin dynamics in both localized and itinerant carriers in Fe/GaAs heterostructures. We find that for excitation densities that push the effective Fermi energy above the mobility edge there exist two distinct precession frequencies in the observed spin dynamics, allowing us to simultaneously monitor both localized and itinerant states. For low magnetic fields (below 3 T) the beat frequency between these two excitations evolves linearly, indicating that the nuclear polarization is saturated almost immediately and that the hyperfine coupling to these two states is the same to within 34%. At higher magnetic fields (above 3 T) the Zeeman energy drives reentrant localization of the photocarriers. Subtracting the constant hyperfine contribution from both sets of data allows us to extract the Lande g-factor for each state and estimate their energy relative to the bottom of the conduction band, yielding -2.16 meV and 17 meV for localized and itinerant states, respectively. This work advances our fundamental understanding of spin-spin interactions between electron and nuclear spin species, as well as between localized and itinerant electronics states, and therefore has implications for future work in both spintronics and quantum information/computation.

**Acknowledgements**

Primary funding for this research was provided by the Center for Emergent Materials: an NSF MRSEC under award number DMR-1420451. This research is also based upon support for Ezekiel Johnston-Halperin and materials and supplies provided by the U.S. Department of Energy, Office of Science, Office of Basic Energy Sciences, under Award Number DE-SC0001304. The authors acknowledge Bernd Beschoten and Paul Crowell for fruitful scientific discussions.

† To whom correspondence should be addressed. Email: ejh@mps.ohio-state.edu

**FIG. 1.** Schematic density of states diagrams of GaAs and illustrations of state filling under (a) low pump excitation and (b) high pump excitation (see main text for details). Measured Faraday rotation angle ($\theta_{FR}$) vs $\Delta t$ for a Fe/MgO/GaAs heterostructure at $B_{app}=$ 5 T and at $I=$ (c) 48 W/cm$^2$ and (d) 241 W/cm$^2$ and the fitting curves by eq. (1). (e) FFT spectra at $I=$ 48 W/cm$^2$ (black) and $I=$ 241 W/cm$^2$ (blue). The arrows label the characteristic frequencies appearing in (c) and (d). (f) The same TRFR time trace as in Fig. 1 (d) and the fitting curve by eq. (2). The insets in (c), (d) and (f) show the zoom-in TRFR time traces from $\Delta t=$ 0.5 ns to 1.2 ns for clarity.

**FIG. 2.** Larmor frequency ($f$) as a function of $B_{app}$ for both high-frequency ($f_H$, solid square) and low-frequency ($f_L$, open square) components at (a) $I=$ 48 W/cm$^2$, (b) $I=$ 108 W/cm$^2$, (c) $I=$ 241 W/cm$^2$ and (d) $I=$ 455 W/cm$^2$. FFT spectra as a function of $B_{app}$ at (e) $I=$ 48 W/cm$^2$, (f) $I=$ 108 W/cm$^2$ and (g) $I=$ 241 W/cm$^2$ and (h) $I=$ 455 W/cm$^2$.

**FIG. 3.** (a) Larmor frequency ($f$) as a function of pump power density ($I$) for both high-frequency ($f_H$, solid square) and low-frequency ($f_L$, open square) components at $B_{app}=$ 5T. (b) Beat frequency ($f_{beat}$) as a function of $B_{app}$ for $I=$ 455 W/cm$^2$, 241 W/cm$^2$ and 108 W/cm$^2$. (c) Power density dependence of $f/B_{app}$ and effective g factors for $f_H$ and $f_L$ components.

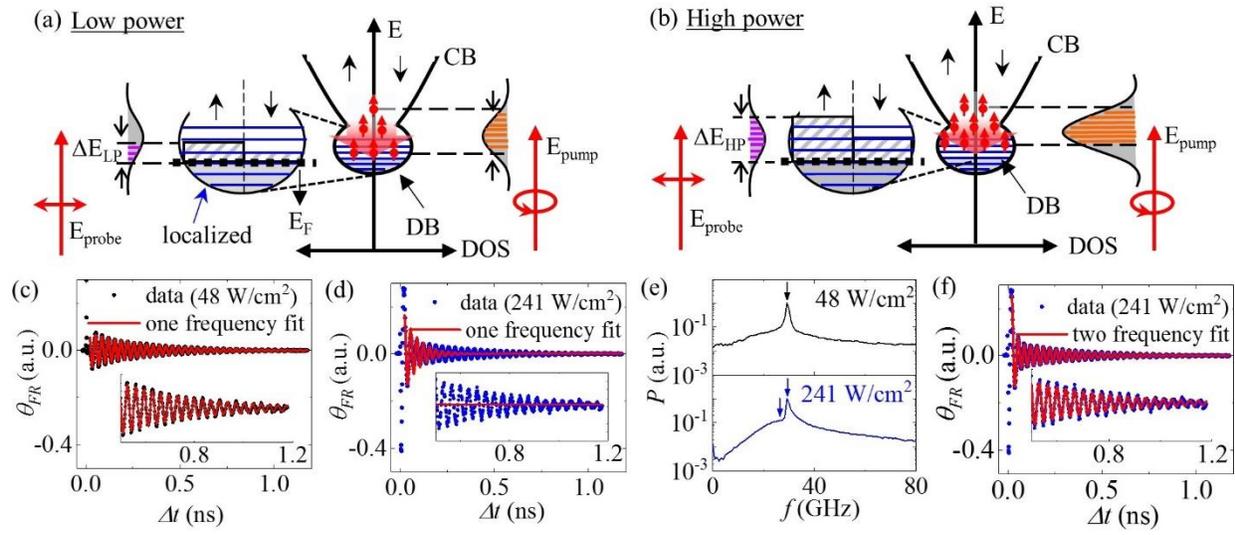

**Figure 1** Yu-Sheng Ou

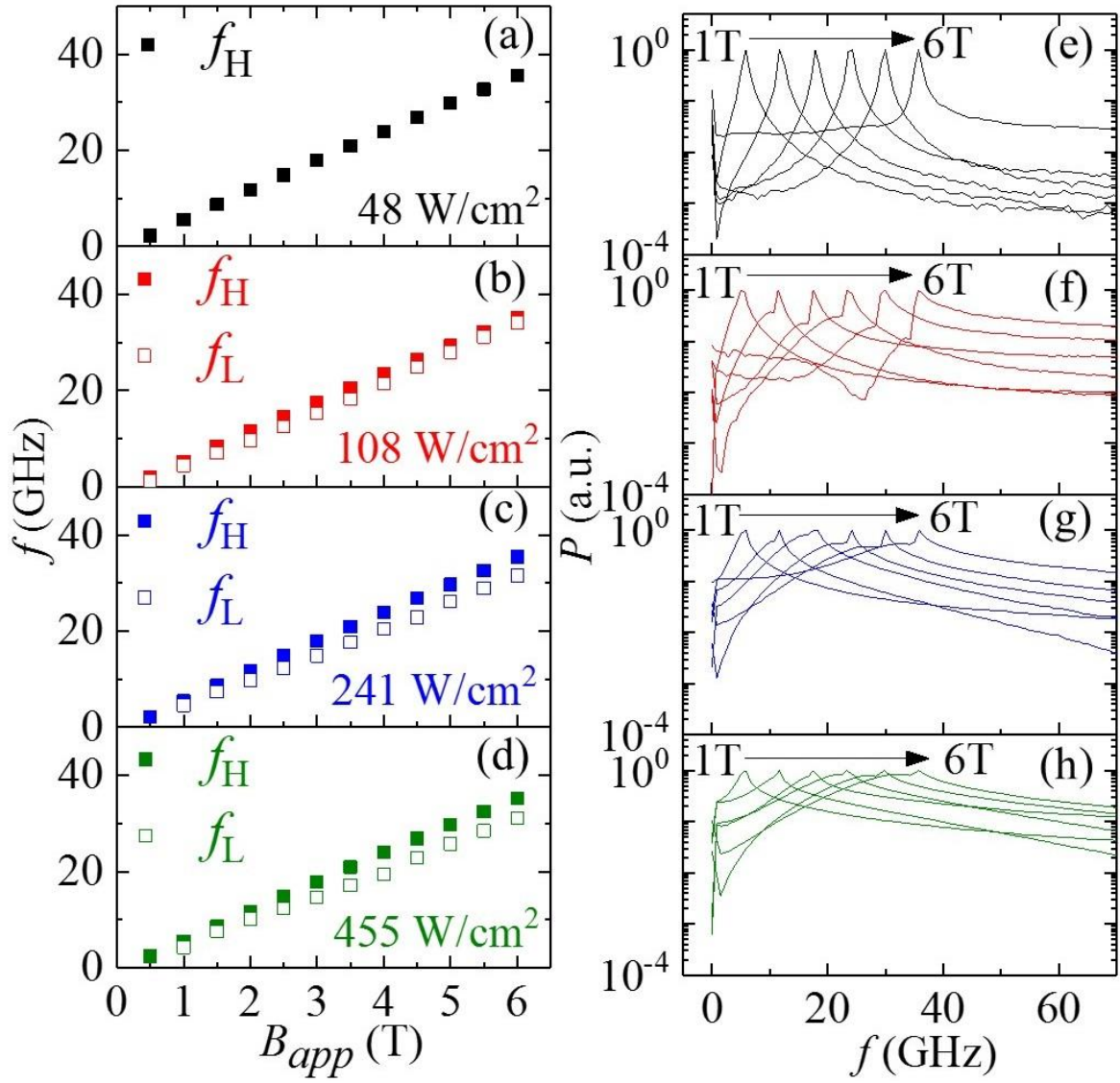

**Figure 2** Yu-Sheng Ou

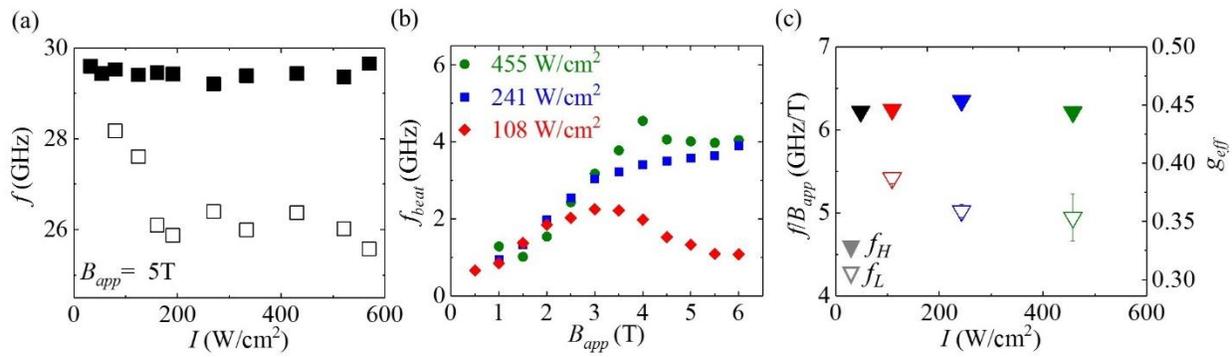

**Figure 3** Yu-Sheng Ou